\documentclass[conference]{IEEEtran}
\IEEEoverridecommandlockouts
\usepackage{textcomp}
\usepackage[nocompress]{cite}

\usepackage{epsfig, epstopdf, graphicx, balance}
\usepackage{amsmath,amssymb, mathrsfs, amsfonts}
\usepackage[mathscr]{euscript}
\usepackage{microtype} 
\usepackage{tabulary} 

\usepackage{enumitem}
\setlist[itemize]{itemsep=3pt,topsep=0pt,parsep=0pt,partopsep=3pt,leftmargin=2em}
\setlist[enumerate]{itemsep=3pt,topsep=0pt,parsep=0pt,partopsep=3pt,leftmargin=2em}

\usepackage[hidelinks]{hyperref}

\usepackage[scaled=0.85]{beramono}

\usepackage{amsthm}
\theoremstyle{definition}
\newtheorem{defn}{Definition} 
\theoremstyle{plain}

\usepackage{mathtools}

\usepackage[linesnumbered,ruled,vlined]{algorithm2e}
\SetAlgorithmName{Algorithm}{Algorithm}{Algorithm}
\IncMargin{0.5em}
\SetCommentSty{textnormal}
\SetNlSty{}{}{:}
\SetAlgoNlRelativeSize{0}
\SetKwInput{KwGlobal}{Global}
\SetKwInput{KwPrecondition}{Precondition}
\SetKwProg{Proc}{Procedure}{:}{}
\SetKwProg{Func}{Function}{:}{}
\SetKw{And}{and}
\SetKw{Or}{or}
\SetKw{To}{to}
\SetKw{DownTo}{downto}
\SetKw{Break}{break}
\SetKw{Continue}{continue}
\SetKw{SuchThat}{\textit{s.t.}}
\SetKw{WithRespectTo}{\textit{wrt}}
\SetKw{Iff}{\textit{iff.}}
\SetKw{MaxOf}{\textit{max of}}
\SetKw{MinOf}{\textit{min of}}
\SetKwBlock{Match}{match}{}{}

\usepackage{array}
\newcolumntype{L}[1]{>{\raggedright\let\newline\\\arraybackslash\hspace{0pt}}m{#1}}
\newcolumntype{C}[1]{>{\centering\let\newline\\\arraybackslash\hspace{0pt}}m{#1}}
\newcolumntype{R}[1]{>{\raggedleft\let\newline\\\arraybackslash\hspace{0pt}}m{#1}}

\usepackage{color}
\usepackage[usenames,svgnames]{xcolor}
\usepackage{soul}
\soulregister\cite7
\soulregister\ref7
\soulregister\pageref7
\soulregister\autoref7
\soulregister\eqref7

\ifdefined\NoHighlight

\fi

\newcommand*{\PicDir}{.}

\newcommand*{\ScreenDir}{.}
\newcommand*{\Pic}[2]{\PicDir /#2.#1}

\newcommand*{\Screen}[2]{\ScreenDir /#2.#1}

\newcommand{\wordy}[1]{#1}

\renewcommand{\wordy}[1]{}

\newcommand*{\CaptionBeforeVSpace}{\vspace*{0pt}}  
\newcommand*{\FigBottomVSpace}{\vspace*{-2ex}}

\newcommand*{\seq}[3]{#1_{#2}, #1_{2}, \dotsc, #1_{#3}}
\newcommand*{\range}[3]{#1_{#2}\dotsc#1_{#3}}

\newcommand*{\forkbase}{ForkBase}\soulregister\forkbase7
\newcommand*{\webui}{\forkbase{} Web UI}\soulregister\webui7
\newcommand*{\cli}{\forkbase{} CLI}\soulregister\cli7

\newcommand*{\git}{Git}\soulregister\git7
\newcommand*{\github}{GitHub}\soulregister\github7
\newcommand*{\gitlab}{GitLab}\soulregister\gitlab7
\newcommand*{\gitflow}{Gitflow}\soulregister\gitflow7
\newcommand*{\dropbox}{Dropbox}\soulregister\dropbox7

\newcommand*{\siri}{SIRI}
\newcommand*{\postree}{POS-Tree}
\newcommand*{\bplustree}{B$^{+}$-tree}
\newcommand*{\fnode}{\textit{FNode}}
\newcommand*{\uid}{\textit{uid}}

\begin{document}

\title{ForkBase: Immutable, Tamper-evident Storage Substrate for Branchable Applications}

\newcommand{\AuthNameSep}{,\ }  
\newcommand{\AuthNameVSep}{\vspace{.5ex}}  %
\newcommand{\AuthNameVSpace}{\vspace{1.2ex}}
\newcommand{\AuthAffSep}{\hspace{1.5em}}
\newcommand{\AuthAffVSep}{}  
\newcommand{\AuthAffVSpace}{\vspace{.8ex}}
\newcommand{\AuthEmailSep}{,\ \ }  
\newcommand{\AuthEmailVSep}{\vspace{.2ex}}

\newcommand*{\AuthName}[2]{#2$^{#1}$}
\newcommand*{\AuthNameWithNote}[3]{\AuthName{#1}{#2}\titlenote{#3}}
\newcommand*{\AuthAff}[2]{$^{#1}${#2}}
\newcommand*{\AuthEmail}[2]{$^{#1}${#2}}
\newcommand*{\AuthEmails}[3]{$^{#1}${\{#3\}@#2}}

\author{
  \IEEEauthorblockN{
    \AuthName{1}{Qian Lin}\AuthNameSep
    \AuthName{1}{Kaiyuan Yang}\AuthNameSep
    \AuthName{2}{Tien Tuan Anh Dinh}\AuthNameSep
    \AuthName{6}{Qingchao Cai}\AuthNameSep
    \AuthName{3}{Gang Chen}\AuthNameSep
    \AuthName{1}{Beng Chin Ooi}\AuthNameSep
  }\IEEEauthorblockN{  
    \AuthName{1}{Pingcheng Ruan}\AuthNameSep
    \AuthName{5}{Sheng Wang}\AuthNameSep
    \AuthName{1}{Zhongle Xie}\AuthNameSep
    \AuthName{4}{Meihui Zhang}\AuthNameSep
    \AuthName{7}{Olafs Vandans}
  }\AuthNameVSpace
  \IEEEauthorblockA{
    \AuthAff{1}{National University of Singapore}\AuthAffSep
    \AuthAff{2}{Singapore University of Technology and Design}\AuthAffSep
    \AuthAff{3}{Zhejiang University}\AuthAffSep
  }\IEEEauthorblockA{  
    \AuthAff{4}{Beijing Institute of Technology}\AuthAffSep
    \AuthAff{5}{Alibaba Group}\AuthAffSep
    \AuthAff{6}{Hudson River Trading Singapore}\AuthAffSep
    \AuthAff{7}{Synspective Inc.}
  }\AuthAffVSpace
  \IEEEauthorblockA{
    \AuthEmails{1}{comp.nus.edu.sg}{linqian, yangky, ooibc, ruanpc, zhongle}\AuthEmailSep
    \AuthEmail{2}{dinhtta@sutd.edu.sg}\AuthEmailSep
    \AuthEmail{3}{cg@zju.edu.cn}\AuthEmailSep
  }\IEEEauthorblockA{  
    \AuthEmail{4}{\href{mailto: meihui_zhang@bit.edu.cn}{meihui\_zhang@bit.edu.cn}}\AuthEmailSep
    \AuthEmail{5}{sh.wang@alibaba-inc.com}\AuthEmailSep
    \AuthEmail{6}{qcai@hudson-trading.com}\AuthEmailSep
    \AuthEmail{7}{to@olafs.eu}
  }
}

\maketitle

\begin{abstract}

Data collaboration activities typically require systematic or protocol-based coordination to be scalable. 
\git{}, an effective enabler for collaborative coding, has been attested for its success in countless projects around the world.  
Hence, applying the \git{} philosophy to general data collaboration beyond coding is motivating. 
We call it \textit{\git{} for data}. 
However, the original \git{} design handles data at the file granule, which is considered too coarse-grained for many database applications. 
We argue that \git{} for data should be co-designed with database systems. 
To this end, we developed \forkbase{} to make \git{} for data practical. 
\forkbase{} is a distributed, immutable storage system designed for data version management and data collaborative operation.
In this demonstration, we show how \forkbase{} can greatly facilitate collaborative data management and how its novel data deduplication technique can improve storage efficiency for archiving massive data versions.

\end{abstract}

\maketitle

\section{Introduction}

Data analytics and machine learning activities generally target at insights from data and further exploit them to enhance applications. 
Processing on the same specific dataset usually involves multiple disciplines that run analytics or data engineering independently. 
And a collaboration is established with respect to the multi-entity efforts towards common goals. 
Efficient data collaboration comes from coordination and storage support.
Most of the existing solutions for collaborative data coordination are built within the application. 
Such ad-hoc approach not only wastes development effort that is hardly reusable across different applications, but also misses the opportunity to optimize the underlying storage. 
Therefore, collaboration-oriented data management must rely on systematic or protocol-based coordination for scaling. 


\git{} has been one of the most productive solutions in the practice of collaborative code management.
This has been attested by its widespread adoption in countless projects around the world.
In the context of database systems, many data collaboration applications have similar collaborative coordination requirements that 
\git{} could potentially fulfill. 
Hence, bringing the \git{} semantics into database management would benefit many kinds of branchable data applications. 
We call such integration \textit{\git{} for data}.
\autoref{tab:compare} briefly compares state-of-the-art \git{} for data systems.

\begin{table*}[htbp]
\caption{Comparison with Related Data Versioning Systems}
\begin{center}
\begin{tabular}{|c|c|c|c|c|}
\hline
  \textbf{\textbf{System}} & \textbf{\textit{Data Model}} & \textbf{\textit{Deduplication}} & \textbf{\textit{Tamper Evidence}} & \textbf{\textit{Branching}} \\
  \hline
  {\forkbase{}~\cite{vldb18:Wang}} & {structured/unstructured, immutable} & {page level} & {root hash of Merkle DAG} & {\git{}-like} \\
  \hline
  {DataHub~\cite{Anant_DataHub} \& Decibel~\cite{vldb16:Maddox}} & {structured (table), mutable} & {table oriented} & {none} & {ad-hoc} \\
  \hline
  {OrpheusDB~\cite{vldb17:Huang}} & {structured (table), mutable} & {table oriented} & {none} & {ad-hoc} \\
  \hline
  {MusaeusDB~\cite{Schule_MusaeusDB}} & {structured (table), mutable} & {table oriented} & {none} & {none} \\
  \hline
  {RStore~\cite{Souvik_RStore}} & {unstructured, mutable} & {key-value} & {none} & {ad-hoc} \\
  \hline
\end{tabular}
\label{tab:compare}
\FigBottomVSpace
\end{center}
\end{table*}


Towards practical \git{} for data, we developed \forkbase{}~\cite{vldb18:Wang} which is collaboration-centric by design and facilitates data forking, data version management and access control for multi-tenant data analytics. 
\forkbase{} provides \git{}-like operations, but focuses on data, its security, immutability and provenance.
Like \git{}, data in \forkbase{} is multi-versioned, and each version uniquely identifies the data content and its history. 


\section{System Overview}
\label{sec:design}

\forkbase{} is a distributed storage system designed for data version management and data collaborative operation.
It internally implements \git{}-compatible data version control and branch management based on Merkle directed acyclic graph (DAG), which empowers tamper evidence and efficient tracking of data provenance.
Moreover, \forkbase{} is endowed with a novel content-based data deduplication technology that can remarkably reduce data redundancy between different data versions in the physical storage as well as efficiently support fast differential queries between data versions.

\begin{figure}[!t]
  \centering
  \includegraphics[width=.9\linewidth]{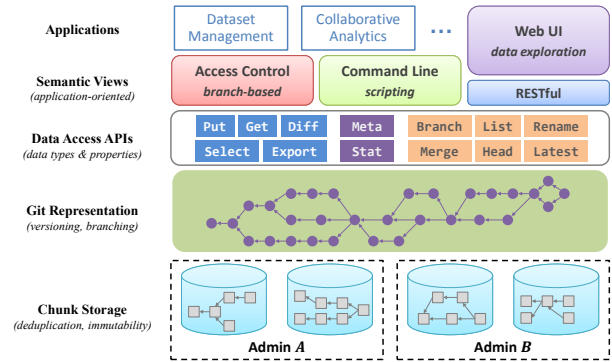}
  \CaptionBeforeVSpace
  \caption{Overview of the \forkbase{} architecture.}
  \label{fig:arch}
  \FigBottomVSpace
\end{figure}

The overall architecture of \forkbase{} is illustrated in \autoref{fig:arch}. 
At the bottom layer, data are deduplicated at the chunk level. 
At the representation layer, versions and branches are organized based on Merkle DAG.
The API layer exposes interfaces for data manipulation\wordy{ (e.g., \texttt{Put} and \texttt{Get} for point data  access, \texttt{Select} for range query, \texttt{Diff} for differential query)}, data property tracking\wordy{ (e.g., \texttt{Meta} for retrieving data object associated meta-data, \texttt{Stat} for probing system status)} and version/branch management\wordy{ (e.g., \texttt{Branch} and \texttt{Merge} for branch management, \texttt{Head} and \texttt{Latest} for version tracking)}. 
Supported data types include primitives (string, number, boolean), blob, map, set and list, as well as composite data structures built on them (e.g., relational table).
The semantic view layer includes access control, command-line interface and RESTful APIs to support applications at the top layer, such as a web-based UI for data exploration across versions and branches. 

The aforementioned key features of \forkbase{} are fundamentally accredited to its novel content-addressable indexing technique. 
In the following, we first elaborate on the design of index, and then describe how it empowers \forkbase{} to achieve effective data deduplication and fast differential query between data versions, as well as tamper evidence. 

\subsection{POS-Tree}

Existing primary indexes in databases focus on improving read and write performance. 
They do not consider data page sharing, which makes page-level deduplication ineffective. 
\forkbase{} addresses this issue by introducing the Pattern-Oriented-Split Tree (\postree{}). 
\postree{} implements the Structurally-Invariant Reusable Index (\siri{})~\cite{vldb18:Wang} which facilitates page sharing among different index instances.

\begin{defn}{\textbf{\siri{}}}
Let $\mathcal{I}$ be an index structure. 
An instance $I$ of $\mathcal{I}$ stores a set of records $R(I) = \{\seq{r}{1}{n}\}$. 
The internal structure of $I$ consists of a collection of pages (i.e.\ index and data pages) $P(I) = \{\seq{p}{1}{m}\}$. 
Two pages are equal if they have identical content and hence can be shared.
$\mathcal{I}$ is called an instance of \siri{} if it has the following properties: 
\begin{enumerate}
\item[(1)]
Structurally invariant: For any instance $I_{1}, I_{2} \in \mathcal{I}$, 
\begin{equation}\label{siri:struct_invariant}
R(I_{1}) = R(I_{2}) \Longleftrightarrow P(I_{1}) = P(I_{2})
\end{equation}

\item[(2)]
Recursively identical: For any instance $I_{1}, I_{2} \in \mathcal{I}$ such that $R(I_{2}) = R(I_{1}) + \{r\}$ for any record $r \not\in I_{1}$, 
\begin{equation}\label{siri:recursive_id}
|P(I_{2}) - P(I_{1})| \ll |P(I_{2}) \cap P(I_{1})|
\end{equation}

\item[(3)]
Universally reusable: For any instance $I_{1} \in \mathcal{I}$ and page $p \in P(I_{1})$, there exists another instance $I_{2} \in \mathcal{I}$ such that
\begin{equation}\label{siri:universe_reuse}
(|P(I_{2})| > |P(I_{1})|) \wedge (p \in P(I_{2}))
\end{equation}
\end{enumerate}
\end{defn}

Specifically, Property~\eqref{siri:struct_invariant} means that the internal structure of an index instance is uniquely determined by the set of records. 
By avoiding the structural variance caused by the order of modifications, all pages between two logically identical index instances can be pairwisely shared. 
Property~\eqref{siri:recursive_id} means that an index instance can be represented recursively by smaller instances with little overhead, while the third property ensures that a page can be reused by many index instances. 
By avoiding the structural variance caused by index cardinalities, Property~\eqref{siri:universe_reuse} means that a large index instance can reuse pages from smaller instances. 
As a result, instances with overlapping content can share a large portion of their sub-structures.

\postree{} being an instance of \siri{} inherits the above properties. 
Moreover, \postree{} is additionally endowed with the following three properties: it is a probabilistically balanced search tree; it is efficient to find differences and to merge two instances; and it is tamper evident.
This structure 
resembles a combination of a \bplustree{} and a Merkle tree~\cite{crypto87:Merkle}. 
In \postree{}, the node (i.e.\ page) boundary is defined as patterns detected from the contained entries, which avoids structural differences. 
Specifically, to construct a node, we scan the target entries until a pre-defined pattern occurs, and then create a new node to hold the scanned entries. 

\begin{figure}[!t]
  \centering
  \includegraphics[width=.85\linewidth]{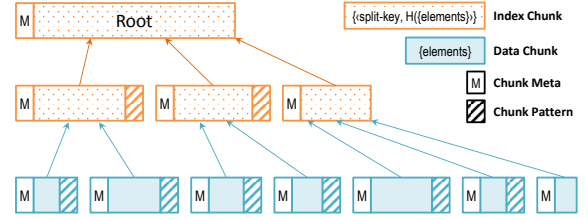}
  \CaptionBeforeVSpace
  \caption{Pattern-Oriented-Split Tree (\postree{}).}
  \label{fig:pos_tree}
  \FigBottomVSpace
\end{figure}


\autoref{fig:pos_tree} illustrates the structure of a \postree{}. 
Each node in the tree is stored as a page, which is the unit for deduplication. 
The node is terminated with a detected pattern, unless it is the last node of a certain level.
Similar to a \bplustree{}, an index node contains one entry for each child node. 
Each entry consists of a child node's identifier and the corresponding split key. 
To look up a specific key, we adopt the same strategy as in the \bplustree{}, i.e., following the path guided by the split keys. 
\postree{} is also a Merkle tree in the sense that the child node's identifier is the cryptographic hash value of the child (e.g., derived from SHA-256 hash function) instead of memory or file pointers. 
The mapping from the node identifier to storage pointer is maintained externally.


In order to avoid structural variance for \postree{} nodes, we define patterns similar to content-based slicing~\cite{sosp01:Muthitacharoen} used in file deduplication systems. 
These patterns help split the nodes into smaller sizes on average. 
Given a $k$-byte sequence $(\seq{b}{1}{k})$, let $\Phi$ be a function taking $k$ bytes as input and returning a pseudo-random integer of at least $q$ bits. 
The pattern occurs if and only if: 
\[
\Phi(\seq{b}{1}{k})\;\text{MOD}\;2^{q} = 0
\]
In other words, the pattern occurs when the function $\Phi$ returns $0$ for the $q$ least significant bits. 
This pattern can be implemented via rolling hashes which support continuous computation over sequence windows and offer satisfactory randomness.
In particular, \postree{} applies the \textit{cyclic polynomial hash}, which is of the form: 
\[
\Phi(\range{b}{1}{k}) = \delta(\Phi(\range{b}{0}{k-1})) \oplus \delta^{k}(\Gamma(b_{0})) \oplus \delta^{0}(\Gamma(b_{k}))
\]
where $\oplus$ is the exclusive-or operator, and $\Gamma$ maps a byte to an integer in $[0, 2^{q})$. 
$\delta$ is a function that shifts its input by $1$ bit to the left, and then pushes the $q$-th bit back to the lowest position.
As a consequence, for each iteration the above recursion removes the oldest byte and adds the new latest one. 

Initially, the entire list of data entries is treated as a byte sequence, and the pattern detection process scans it from the beginning.
When a pattern occurs, a node is created from recently scanned bytes. 
If a pattern occurs in the middle of an entry, the page boundary is extended to cover the whole entry, so that no entries are stored across multiple pages. 
In this way, each node (except for the last node) ends with a pattern.

\begin{figure}[!t]
  \centering
  \includegraphics[width=.78\linewidth]{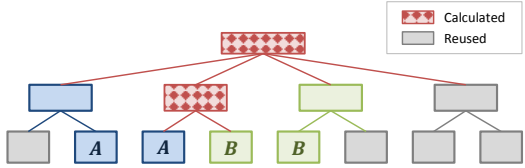}
  \CaptionBeforeVSpace
  \caption{Three-way merge of two \postree{}s reuses disjointly modified sub-trees to build the merged tree.}
  \label{fig:merge}
  \FigBottomVSpace
\end{figure}

\subsection{Diff and Merge}

\postree{} supports fast \texttt{Diff} operation which identifies the differences between two POS-Tree instances. 
Because two sub-trees with identical content must have the same root id, the \texttt{Diff} operation can be performed recursively by following the sub-trees with different ids, and pruning ones with the same ids. 
The complexity of \texttt{Diff} is therefore $O(D log(N))$, where $D$ is the number of different leaf nodes and $N$ is the total number of data entries.

\postree{} supports three-way merge which consists of a diff phase and a merge phase. 
In the diff phase, two objects $A$ and $B$ are diffed against a common base object $C$, which results in $\Delta_{A}$ and $\Delta_{B}$ respectively. 
In the merge phase, the differences are applied to one of the two objects, i.e., $\Delta_{A}$ is applied to $B$ or $\Delta_{B}$ is applied to $A$.
In conventional approaches, the two phases are performed element-wise. 
In \postree{}, both phases can be done efficiently at sub-tree level. 
More specifically, we do not need to reach leaf nodes during the diff phase, as the merge phase can be performed directly on the largest disjoint sub-trees that cover the differences, instead of on
individual leaf nodes, as illustrated in \autoref{fig:merge}.

\begin{figure}
  \centering
  \includegraphics[width=.8\linewidth]{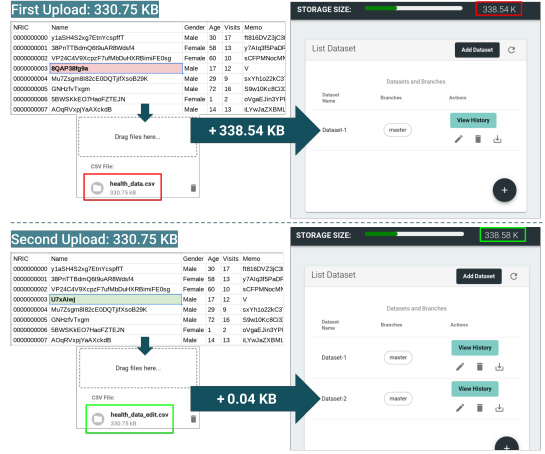}
  \CaptionBeforeVSpace
  \caption{Fine-grained data deduplication in \forkbase{}.}
  \label{fig:dedup}
  \FigBottomVSpace
\end{figure}

\subsection{Storage Efficiency}

Another advantage brought by \postree{} is its powerful data deduplication ability. 
As illustrated in \autoref{fig:pos_tree}, data are split into chunks, each of which is immutable after complete construction and uniquely identified by its SHA-256 hash.
Chunks are materialized into the key-value based physical storage so that each distinct chunk is stored exactly once and can be shared across different data objects according to the identical content, e.g., in the example shown in \autoref{fig:merge}. 


\subsection{Tamper-evident Version}

\forkbase{} adopts an extended key-value data model: each object is identified by a key, and contains a value of a specific type. 
A key may have multiple branches. 
Given a key we can retrieve not only the current value in each branch, but also its historical versions.
Similar to other data versioning systems, \forkbase{} organizes versions in a directed acyclic graph (DAG) called \textit{version derivation graph}: Each node in the graph is a structure called \fnode{}, and links between \fnode{}s represent their derivation relationships. 
Each \fnode{} is associated with a \uid{} representing its version, which can be used to retrieve the value. 
The \uid{} uniquely identifies both the object value and its derivation history, based on the content stored in the \fnode{}. 
Two \fnode{}s are considered equivalent, i.e., having the same \uid{}, when they have both the same value and derivation history. 
This is due to the use of \postree{}---a structurally invariant Merkle tree---to store the values. 
In addition, the derivation history is essentially a hash chain formed by linking the bases fields, thus two equal \fnode{}s must have the same history.

Each \uid{} is tamper evident. 
Given a \uid{}, the user can verify the content and history of the returned \fnode{}. 
This integrity property is guaranteed under the following threat model: the storage is malicious, but the users keep track of the latest $uid$ of every branch that has been committed. 
Instead of introducing a new tamper evidence design, \forkbase{} supports this property efficiently as a direct benefit from the \postree{} design.

\section{Demonstration}
\label{sec:demo}

Our demonstration using the \webui{}
aims to highlight some novel aspects of the \forkbase{} system. 
In particular, we focus on the capability of \forkbase{} to deduplicate data, perform fast differential query, branch dataset with \git{}-compatible semantics, and track versions with tamper evidence.

\subsection{Data Deduplication}

First, we showcase the fine-grained data deduplication feature of \forkbase{}.
As shown in \autoref{fig:dedup}, two external CSV datasets with a single-word difference in terms of text content are loaded into \forkbase{} as two separate datasets. 
Loading the first dataset increases $338.54$~KB to the storage, but afterwards loading the second dataset only increases $0.04$~KB. 
As the two datasets share a large portion of duplicated data, \forkbase{} can effectively detect such redundancy and consequently store only the marginal difference into the underlying immutable storage for the second loading. 
This manifests 
that \forkbase{} can significantly improve storage efficiency through its fine-grained data deduplication.

\subsection{Fast Differential Query}

\begin{figure}
  \centering
  \includegraphics[width=.8\linewidth]{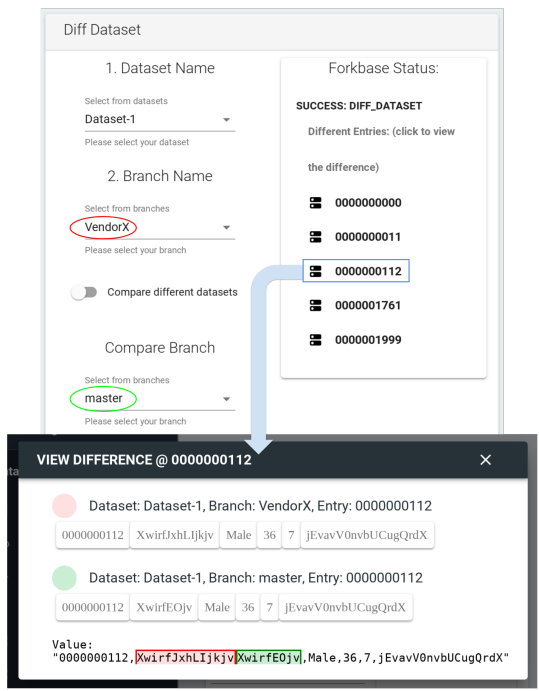}
  \CaptionBeforeVSpace
  \caption{An example of differential query in \forkbase{}.}
  \label{fig:diff}
  \FigBottomVSpace
\end{figure}

\forkbase{} is able to perform fast differential query to retrieve differences among data versions and branches. 
To showcase such feature, we designed the \webui{} to visualize differences between datasets and their branches.
For example, \autoref{fig:diff} shows the result of performing \texttt{Diff} operation between the \texttt{master} and \texttt{VendorX} branches of \texttt{Dataset-1}.
Data differences are highlighted at multiple scopes, e.g., from dataset to data entry. 
This is akin to the \git{}-diff utility which helps user identify the changes of data. 

\subsection{Tamper Evidence and Validation}

\begin{figure}
  \centering
  \includegraphics[width=.8\linewidth]{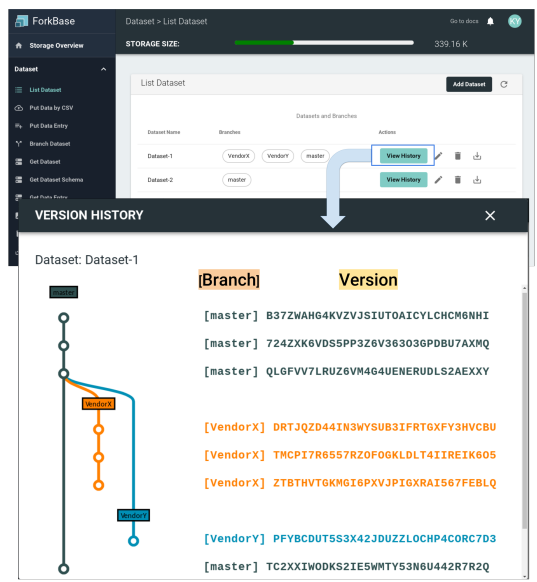}
  \CaptionBeforeVSpace
  \caption{Versioning for validation and tamper evidence.}
  \label{fig:history}
  \FigBottomVSpace
\end{figure}

Along with data updates, each \texttt{Put} operation is stamped with a unique version that is appended to the corresponding branch of the dataset, as shown in \autoref{fig:history}. 
Data versions in \forkbase{} are generated according to the Merkle root hash of data chunks~\cite{crypto87:Merkle, vldb18:Wang}, and encoded using the RFC 4648 Base32 alphabet~\cite{RFC4648}.
Such versioning scheme enables \forkbase{} to provide tamper evidence against malicious storage providers. 
Given a version, the application can fetch the corresponding data from the storage provider (i.e., physical storage) and 
validate the content and its history by checking whether the Merkle root hash calculated on the spot is identical to the data version. 
This guarantees data stored in \forkbase{} is tamper-proof in spite of the underlying storage infrastructure.

\section{Conclusion}
\label{sec:conclude}

This demonstration sheds light on the basic workflow of \git{}-like data management for collaborative data processing.
By pushing down the \git{}-compliant versioning and branching semantics from the application layer to the storage layer, \forkbase{} benefits various kinds of branchable applications built on top of it with reduced development effort. 
The fine-grained data deduplication feature of \forkbase{} favors improved storage efficiency in collaborative data activities.

\section*{Acknowledgment}

This work is supported by Singapore Ministry of Education (MOE) Academic Research Fund Tier 3 under MOE's official grant number MOE2017-T3-1-007.

\balance

\newcommand{\BIBdecl}{\setlength{\itemsep}{0.48em}} 
\bibliographystyle{IEEEtran}
\bibliography{references-trim}

\end{document}